\documentclass[iop,revtex4]{emulateapj}
\usepackage{amssymb,stackengine}
\usepackage{xcolor}
\definecolor{rkka}{RGB}{219,66,32}

\begin{document}

\newcommand{\kms}{km~s$^{-1}$\,}
\newcommand{\msun}{$M_\odot$\,}
\newcommand{\eb}{\begin{equation}}
\newcommand{\ee}{\end{equation}}

\renewcommand{\topfraction}{1.0}
\renewcommand{\bottomfraction}{1.0}
\renewcommand{\textfraction}{0.0}

\title{Spin-orbit resonances of high-eccentricity asteroids: regular, switching, and jumping}
\shorttitle{Spin resonances of high-eccentricity asteroids}

\author{Valeri V. Makarov}
\affil{U.S. Naval Observatory, 3450 Massachusetts Ave., Washington, DC 20392-5420, USA}
\email{valeri.makarov@gmail.com}

\author{Alexey Goldin}
\affil{Teza Technology, 150 N Michigan Ave, Chicago IL 60601, USA}
\email{alexey.goldin@gmail.com}

\author{Dimitri Veras}
\affil{Centre for Exoplanets and Habitability, University of Warwick, Coventry CV4 7AL, UK\\
Department of Physics, University of Warwick, Coventry CV4 7AL, UK}
\email{d.veras@warwick.ac.uk}
\thanks{STFC Ernest Rutherford Fellow}

\begin{abstract}
Few solar system asteroids and comets are found in high eccentricity orbits ($e > 0.9$) but in the primordial
planetesimal disks and in exoplanet systems around dying stars such objects are believed to be common.
For 2006 HY51, the main belt asteroid with the highest known eccentricity 0.9684, we investigate the
probable rotational states today using our computer-efficient chaotic process simulation method. Starting with random
initial conditions, we find that this asteroid is inevitably captured into stable spin-orbit resonances
typically within tens to a hundred Myr. The resonances are confirmed by direct integration of the equation
of motion in the vicinity of end-points. Most resonances are located at high spin values above 960 times 
the mean motion (such as 964:1 or 4169:4), corresponding
to rotation periods of a few days. We discover three types of resonance in the high-eccentricity
regime: 1) regular circulation with weakly librating aphelion velocities and integer-number
spin-orbit commensurabilities; 2) switching resonances of higher order with orientation alternating between
aligned (0 or $\pi$) and sidewise ($\pi/2$) angles at aphelia and perihelia; 3) jumping resonances with aphelion
spin alternating between two quantum states in the absence of spin-orbit commensurability. The islands of
equilibrium are numerous at high spin rates but small in parameter space area, so that it takes millions of
orbits of chaotic wandering to accidentally entrap in one of them. We discuss the implications of this discovery
for the origins and destiny of high-eccentricity objects and the prospects of extending this analysis to the
full 3D treatment.
\end{abstract}



\section{Introduction}
\label{sec:intro}

Some asteroids in the Solar system have eccentricities above 0.95. Their origin and 
destiny are not clear. They may be remnants of the primordial asteroid belt, results 
of relatively recent interaction with planets, or even captured interstellar objects. At such high eccentricities,
triaxial celestial bodies can acquire very high prorate spin rates over billions of years of chaotic
evolution \citep{mav} resulting in a rotational break-up and destruction.  The most notable 
high-eccentricity object in the JPL Horizons database is 2006 HY51 with an eccentricity 
of 0.9684\footnote{\url{https://ssd.jpl.nasa.gov/sbdb.cgi\#top}}, but a few other objects with slightly smaller 
eccentricity have been detected. These Apollo class asteroids are Near-Earth Objects. Similar 
objects were involved in the bombardment of inner planets and accretion of primordial terrestrial 
planets. Around white dwarf planetary systems, highly eccentric asteroids are thought to be the 
primary progenitor of debris disks \citep{jura2003, debetal2012, veretal2014, malper2020a,
malper2020b, ver+20} and observed metallic pollution in the photospheres of host stars \citep{zucetal2010, 
koeetal2014}.

Minor planets are markedly non-spherical, and many of them, especially the larger ones, can be
well approximated with triaxial ellipsoids. This idealized model is dynamically represented by three
unequal, mutually orthogonal moments of inertia $A$, $B$, and $C$, in increasing order. The values
are not readily available from observations or remote measurements, but can be estimated from the
shape elongation parameters assuming uniform mass density or a certain density profile. There is a tendency for
the dimensionless degree of triaxiality $\sigma=(B-A)/C$ to increase with decreasing size and mass
as we move from terrestrial planets to moons, minor planets, and comets. Potato-like shapes become prevalent
in the domain of comets and asteroids, with $\sigma$ becoming of the order of 0.1 and larger.
The prolate shape and the gradient of gravitational potential from the central body give rise to
a time-variable torque, which depends on the orientation of the asteroid with respect to the perturber. 
The corresponding equations of motion (Euler's equations) comprise a set of three second order nonlinear 
differential equations, which include
the instantaneous direction cosines of the longest axis and the instantaneous spin
rates about all three principal axes \citep{dan}. A 1D analog of this system is
obtained for the planar case of zero obliquity when the principal axis 
of inertia (corresponding to the moment $C$) is always orthogonal to the orbital plane \citep[e.g.,][]{gol}. 

In this paper, we investigate the rotational states of high-eccentricity triaxial objects in the basic 
1D model (i.e., regarding only the spin about the principal axis of inertia) on the example of 2006 HY51, the most eccentric asteroid on a closed orbit in the solar system. With a period of 4.17 yr and semimajor axis
of 2.59 au, this asteroid spends most of the time at great distances from the Sun experiencing vanishingly
small external torque and, hence, rotating practically at a constant rate. The situation changes dramatically
when it flashes through the perihelion at the closest separation 0.082 au, where a short but powerful burst 
of interaction changes its spin
in a shock-like manner. The direction of this pulse depends mostly on the orientation angle at the point
of closest approach, but the amplitude of impulse also depends on the current spin rate. The result is
a strongly chaotic process, which was theoretically deduced even for much lower eccentricities \citep{wis84, wis87}.

The chaotic 1D spin evolution
was investigated in \cite{mav} by direct integration with stiffness switching. In that paper, however, evolution 
was limited to
a timescale which is roughly six orders of magnitude shorter than the current age of the
solar system due to the slow and sequential integration, which cannot be parallelized. A computer-efficient and
fast alternative technique is described in \citep{mak20}, which allows running parallel simulation trials spanning
gigayears. This method takes advantage of the deterministic mapping between the two phase space parameters on the
scale of a single orbit, the orientation angle $\theta$ at the time of aphelion, and the update of rotation velocity
resulting from the perihelion impulse. We used the higher fidelity version of the method, which is computationally
more demanding, involving double interpolation mapping of the aphelion parameter space.

This paper is organized as follows.
In Section \ref{sim.sec}, we briefly describe the computation for 2006 HY51 that led us to the discovery of
multiple stable spin-orbit resonances. These resonances are confirmed by numerical integration with initial
parameters in the vicinity of selected points of equilibrium in Section \ref{map.sec} and parameter space
cross-sections are mapped for some of them. Three different kinds of high-spin resonances are described
in Section \ref{kind.sec}, one of which is not spin-orbit commensurate. 
Conclusions are drawn and possible directions of future research are discussed in Section \ref{con.sec}.

\section{Long-term simulations of rotation}
\label{sim.sec}
\begin{figure}
\plotone{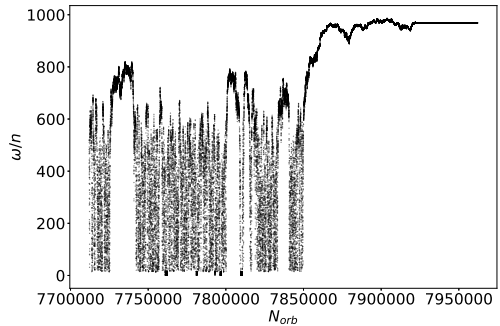}
\caption{Simulated chaotic evolution of the rotation velocity of 2006 HY51 followed by a spontaneous
capture into a commensurate spin-orbit resonance after $7.92\cdot 10^6$ orbits. A small segment of a
$2.5\cdot 10^8$ orbits simulation produced by the fast mapping method is shown.
\label{capt.fig} }
\end{figure}

We made use of the more computer-intensive version of the fast tuple generation method described in
\citep{mak20}\footnote{A julia script implementing this simulation is openly available at 
\url{https://github.com/agoldin/2006HY51}}. The basic idea is to replace the costly ODE integration with 
generation of $\{d_{\omega, i}, \theta_{i+1}\}$ 
tuples for perihelion and aphelion times from precomputed 2D interpolation functions, which turn out to be smooth and well-behaved in the domain of interest. In this paper, $\theta$ is the orientation angle of the  
ellipsoidal asteroid's longest
axis in the orbital plane, i.e., the angle between the axis of the smallest moment of inertia
$\cal{A}$ and the fixed line of apses. It's time derivative, $\omega=\dot\theta$, is the sidereal rotation velocity,
and $d_\omega$ is the change of $\omega$ between two consecutive aphelia. As in the original paper,
the motion is confined to the orbital plane, the gravitational torque is always orthogonal to the orbital plane, and
only the force from the Sun is considered, neglecting small perturbations from the planets or the YORP effect.
Possible effects of YORP and external perturbations are briefly discussed in \S \ref{map.sec}.
Our main computation included 512 separate and independent
random-seeded trials for $2.5\cdot 10^8$ orbits each, i.e., each was longer than 1 Gyr, assuming a triaxiality
parameter $\sigma=(B-A)/C=0.2$. It confirmed that 2006 HY51 could not come close to the rate of several thousand $n$ required for rotational break-up, by roughly an order of magnitude.
The unexpected result was that all 512 simulations ended up in equilibrium states (resonances) within the simulation time 
span. A long-term or indefinitely long equilibrium state characterized by an aphelion spin rate
varying within a finite range despite the powerful perturbations at perihelion passages, as opposed to rapidly changing
chaotic rotation, is called a spin-orbit
resonance in this paper. As we will see in the following, this equilibrium rotation state can be achieved even
without an integer number commensurability of the spin rate and orbital frequency. Numerically, once a state of this kind was achieved, 
our fast tuple-generating simulation stopped behaving chaotically and 
remained within certain narrow ranges of phase space parameters. Fig. \ref{capt.fig} shows the aphelion spin of a small
segment of one such 
simulation. It depicts the peculiar character of spin evolution, when the random process appears to be bounded
to a certain wide range of prograde velocities (approximately $12 n$ and $1100 n$), 
transversing it relatively rapidly from one end to the other,
and stalling for extended periods of time at the low end. These features are related to the remarkable distribution
of impulses in the parameter space \citep[see Fig. 4 in][]{mav}. Vanishingly small changes of velocity are
possible when the process is cornered at the low bound adjacent to the separatrix. Likewise, although there is no
fixed upper bound, the velocity updates become so small at high prograde velocities that the process may linger
there for longer times. Returning to Fig. \ref{capt.fig}, in one of such episodes, the asteroid suddenly
stopped behaving chaotically and continued to rotate at a nearly constant aphelion rate.

Although the overall character and statistics of these long-term simulations are in excellent
agreement with direct integration trials, the latter did not reveal the existence of resonances. The
reason is that we performed only relatively short ($10^5$ orbits) integrations, which are computationally
heavy. Because of the rapid and powerful variation of the integrated parameters $\theta$ and $\omega$
within a small segment of the orbit around the perihelion, which takes only several days to cross, sufficiently
accurate numerical results can be obtained with stiffness-switching methods, which automatically adjust the time
step to the local gradient of the integrand. 
The process should be run for much longer time spans to see resonance captures. This means that
the probability of random capture is low and it takes millions of orbits to accidentally hit one of
the ``stability islands". Compared to the situation found by \citet{wis84} for solar system moons, we have
a vast sea of chaos with tiny islands of equilibrium interspersed in it at high eccentricity. 

\begin{figure}
\plotone{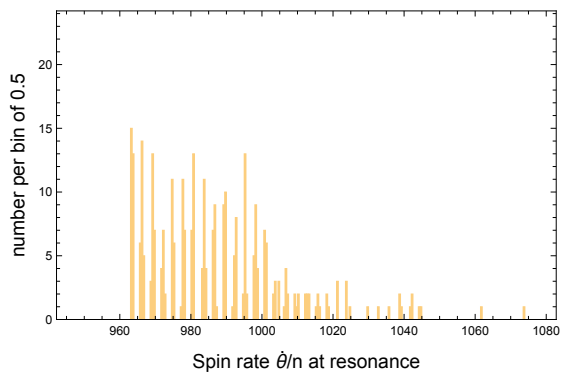}
\caption{Histogram of end spin states of {\it regular} circulation resonances. 
\label{endrate.fig} }
\end{figure}

We can estimate the characteristic time of capture into resonance
by counting the quantiles of the chaotic phase durations. Of the entire set of 512 simulations with random initial
conditions, 25\% were captured within $2.5\cdot 10^6$ orbits, and 50\% -- within $6.5\cdot 10^6$ orbits.
Thus, a significant fraction of trials show resonance capture within 20 to 50 million years. 
The resonance endpoints in $\omega_{\rm ap}/n$ are quantized with the lowest frequency around 963.5
(i.e., a 1927:2 resonance),
except for a small fraction of trials that became stuck at the lower end of the $\omega$ range. The resonance
velocities of the majority high-spin end-points are shown in Fig. \ref{endrate.fig}. We surmise that the
characteristic time of capture into a specific resonance depends on the footprint in the parameter space.

\section{Mapping resonances of 2006 HY51}
\label{map.sec}

To verify the high-spin resonances discovered with the fast-tuple generation method, we performed exact
numerical integration using some of the end-points as initial conditions. One such deeper investigated resonances
corresponds to a peak at the lower end of the distribution in Fig. \ref{endrate.fig}. The spin rate in this 
particular resonance displays a regular circulation behavior varying within a narrow range around $\omega = 966.5 n$. 
The aphelion orientation angle modulo $\pi$ circulates about 0, i.e., the asteroid is aligned with its
longest axis with the direction to the 
Sun. The width of variation depends on the initial perturbation but is limited to the width of the resonance. 
The resonances that are seen as peaks in Fig. \ref{endrate.fig} represent semi-integer spin-orbit commensurabilities
i.e., $\kappa:1$ and $\kappa:2$. 
90\% of our trials that ended in one of the fixed-$\omega$ resonances did this in less than $11\cdot 10^6$ orbits, 
with a median duration of $3.4\cdot 10^6$ orbits.

\begin{figure}
\plotone{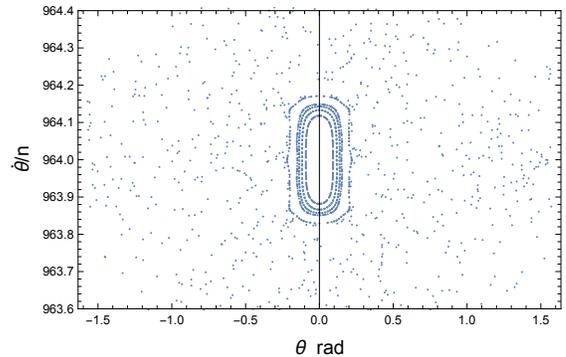}
\caption{Aphelion parameter section of the 964:1 {\it regular} spin-orbit resonance of 2006 HY51 obtained by direct
integration of motion. 
\label{964.fig} }
\end{figure}

Once we have identified approximate location of high spin resonances in the parameter space $\{\theta_{aphelion}$, 
$\omega_{aphelion}\}$ using extensive simulations with the fast tuple-generation method, these tiny zones of equilibrium 
can be mapped in greater detail by direct numerical integration. Fig. \ref{964.fig} shows a parameter space cross 
section (similar to 
a Poincar{\'e} map) at aphelion obtained from 20 integration trials of 200 orbits each with random initial parameters 
in the 
vicinity of the 964:1 resonance. It shows the actual half-width of the circulation island, which is 0.21 rad in 
$\theta$ and 0.17 n in $\omega$.

\begin{figure}
\plotone{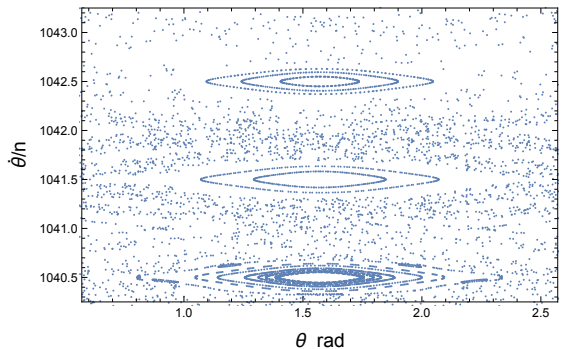}
\caption{Aphelion parameter section of three adjacent spin-orbit resonances of 2006 HY51 with sidewise orientation of the asteroid and half-integer commensurabilities, obtained by direct
integration of motion. 
\label{chain.fig} }
\end{figure}

The islands of stable equilibrium are lined up in the $\{\theta_{aphelion}$, 
$\omega_{aphelion}\}$ parameter space as beads on a string. Multiple resonances are found with $\theta$ modulo $\pi$ close to $\pi/2$, i.e., oriented 
sidewise at aphelia. This is not a new type of spin-orbit resonance since a sidewise capture into resonance has been discussed and deemed possible for the Moon, for example. These fixed-$\omega$ resonances, which we call
regular in this paper, are characterized by a nearly zero net velocity update $d_{\omega}$, because the
asteroid enters the periapse with a $\theta$ modulo $\pi$ close to either 0 (aligned) or $\pi/2$ (sidewise).
A chain of half-integer resonances is mapped in Fig. \ref{chain.fig}. Note that there are multiple resonances
on both side of this sequence outside of the plot.

\begin{figure}
\plotone{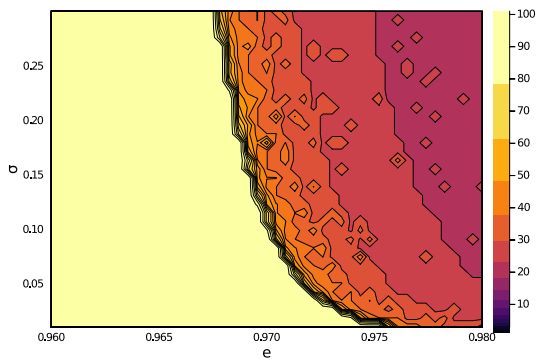}
\caption{Areas of chaos (red colors) and stable spin-orbit resonance (yellow) of 2006 HY51 rotation in the $\sigma$--$e$
parameter space for initial aphelion conditions $\theta(0)=\pi$, $\omega(0)=964\,n$. To distinguish chaotic and resonance
trajectories, the number of orbits is computed (exponentially color-coded in this graph) for the GALI(2) index to fall
below $10^{-12}$, see text. Each integration was limited to 100 orbits.
\label{contour.fig}}
\end{figure}

The cross-section and the very existence of a regular circulation resonance depend on the physical
parameters $\sigma=(B-A)/C$ and eccentricity $e$. We conducted a series of numerical experiments for the stable
and well defined $964:1$ resonance (with the initial spin rate in close vicinity to $\omega=964\,n$).
In the example shown in Fig.~\ref{contour.fig}, we start at the point of resonance with an initial aphelion
angle $\theta=\pi$. To distinguish chaotic trajectories from stable resonance trajectories, we employ the chaos
detection software implemented in Julia by \citet{dat}. The ODE of motion is integrated with two tangent
perturbation vectors pointing from the original point to two random points $\epsilon$
away for 100 orbits. The method uses the GALI(2) alignment index \citep{sko}, which in this 2D case computes the
separation between the two perturbation vectors\footnote{ 
Some explanations how this method works can be found in
\url{https://juliadynamics.github.io/DynamicalSystems.jl/latest/chaos/chaos\_detection/}
}.
As chaotic behavior sets in, the initially orthogonal vectors become increasingly aligned approaching the direction
of the strongest Lyapunov eigenvector. The decay of the area (or distance in this case) between the vectors is
exponential for chaotic trajectories, but it follows a power law for deterministic trajectories. We empirically set
the GALI(2) index threshold at $10^{-12}$. The contour plot shows the number of orbits required for the alignment of 
tangent vectors to drop below the
threshold. The power-law decay for resonance trajectories is too slow to reach the threshold within the 100 orbits,
hence, such trials end up in the saturated yellow-colored area of the graph. Fig.~\ref{contour.fig} shows that
the boundary between chaotic and resonance states in the $\sigma$--$e$ space is very sharp, and this commensurate
spin-orbit resonance exists only if the triaxiality and eccentricity are sufficiently low. With $\sigma=0.2$
assumed for our fast-tuple generation numerical simulations, 2006 HY 51 is located close to the boundary on the
left-hand side, just enough for the stable $964:1$ circulation resonance to emerge.

In the Solar system, as well as in other stellar configurations, asteroids are subject to external
variable forces, or perturbations, apart from the gravitational attraction from the star. The width of resonances
depicted in Figs. \ref{964.fig} and \ref{chain.fig} defines the level of perturbation required to remove the
object from an equilibrium state. E. g., an external acceleration that changes the spin rate by more than $0.17\,n$
on the time scale of one orbit would probably be sufficient to drive 2006 HY 51 out of the 964:1 spin-orbit resonance.
The distribution of main belt asteroid rotation periods is fairly wide and limited on the high end by approximately
2.4 hours \citep{her} for radii greater than approximately 0.1 km with most of the objects spinning quite fast. The
high rates of rotation are commonly attributed to the secular YORP acceleration, which is caused by irregularities
in asteroid's shape, solar irradiation of the surface, and thermal radiation from the surface. Extrapolating
YORP acceleration timescales estimated by \citet{rub}, the expected value for 2006 HY 51 is $\sim 10^6$ yr.
Analytical approximations of the YORP effect as a function of orbital parameters, mass, shape, and stellar luminosity
\citep{sch, vs}, within the great uncertainty of many parameters, provide estimates of the order of $0.1$--$1.0\,n$
per year, which should be enough to remove the asteroid out of a high-spin resonance. Observationally, however,
only much lower accelerations have been detected not exceeding $7\times 10^{-6}$ rad d$^{-1}$ yr$^{-1}$ for Cacus
asteroid \citep{dur}. These estimates refer to objects in smaller orbits with lower eccentricity (up to 0.5).
The YORP effect for extremely high eccentricity is an absolutely unexplored area of research. The solar irradiation
is better approximated with impulse-like bursts at perihelia, and a monotonic acceleration model (spin-up or spin-down)
is not obvious. Furthermore, once the asteroid is captured in a spin-orbit resonance, its orientation is not random
at perihelia where most of the energy deposition takes place. Depending on the specific free libration pattern,
one side of the asteroid may be heated more than the other as long as the resonance lasts. This violates the
starting assumption (which is probably valid for the chaotic stage) that the asteroid absorbs energy equally from
all sides. The condition of anisotropic irradiation may have far-reaching consequences, which are difficult to
guess without detailed modeling and simulation.

A state of resonance may be relatively short-lived for other reasons beside YORP acceleration. Gravitational interactions with
major planets, even of limited magnitude, can change the orbital parameters (eccentricity and pericenter distance,
most importantly) and violate the conditions of resonance. It is reasonable to assume that a sudden change of the
eccentricity, for example, can make a specific resonance at the high end of spin rates unstable, and the object
embarks on another extended chaotic journey. Even more likely, landslides, micrometeorite impacts and other 
stochastic physical processes conspire to regulate the spin, up and down, throughout the evolution.

\section{Three kinds of high-spin resonances}
\label{kind.sec}
Regular circulation resonances constitute approximately 2/3 of our trials and are characterized by the same 
apoastron and periastron $\theta$ modulo $\pi$, which can be close to 0 or $\pi$ (aligned) or 
$\pi/2$ (sidewise), and a stable, weakly librating aphelion rotation velocity\footnote{We call it {\it circulation} resonance to
emphasize the continuously high prograde spin, which is only perturbed with high-amplitude pulses at perihelion
passages. The closest well-known example of circulation resonance is Mercury, albeit with a much slower rotation at
$3:2$ commensurability.}. It is noted that in the aligned apoastron orientation, the geometrically longest
axis of the asteroid is aligned with the direction to the primary attractor, while a sidewise orientation is reached when
the shortest equatorial axis is aligned with that direction. The best known example of the latter in the Solar system
is Mercury, which is sidewise aligned at aphelia. The amplitude of free libration is limited by the width of a particular
resonance. The remaining 1/3 of cases represent two new kinds of spin-orbit resonance, which may not have been described in the literature. 

\begin{figure}
\plotone{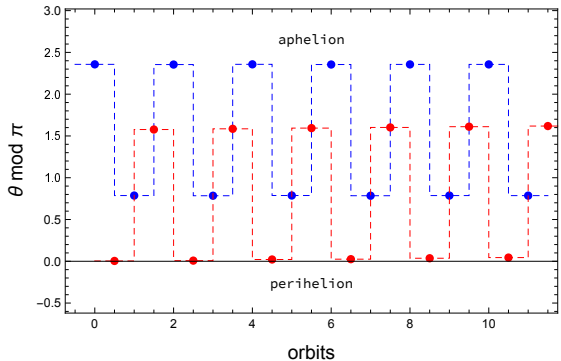}
\caption{Aphelion (blue) and perihelion (red) orientation angles $\theta$ modulo $\pi$ in the 4169:4 {\it switching} spin-orbit resonance of 2006 HY51.
\label{th.fig} }
\end{figure}

The first new kind, called switching resonance, is characterized by a nearly constant aphelion velocity $\omega$, which 
is in a higher order of commensurability $f$ with the mean motion $n$, and aphelion orientation cycling through integer multiples of $\pi/f$. One example is the resonance at $\omega = 1042.25 n$ (i.e., 4169:4, f = 4). The spin rate appears 
to chaotically vary within a very narrow range of $0.0006 n$ around the
mean value, while $\theta_{aphelion}$ modulo $\pi$ switches between 
$\pi/4$ and $3\pi/4$ between each consecutive orbits, see Fig. \ref{th.fig}. The blue dots show
the aphelion normalized orientation angles, while the red dots show the perihelion angles. The dashed lines do not represent
the actual behavior of $\theta$ between the apses but serve only to help the eye to see the switching behavior.
Obviously, this resonance is only possible when the numerator of the spin-orbit commensurability is an odd integer. 
The aphelion tilt of 45 degrees 
with respect to the Sun's direction compensates the fractional part of the relative spin, to the effect that 
the perihelion orientation angle switches between $\pi/2$ (sidewise) and $\pi$ (aligned) in increments of integer multiple of $\pi/2$. Our simulations indicate that this resonance 
is rarely achieved through chaotic evolution, probably due to its narrowness. Although this switching of orientation
may be considered to be a high-order case of the general spin-orbit commensurability, no analogs exist in the
low-eccentricity solar system, with Mercury having the highest order of commensurability 3:2 \citep{noy}, but still not
switching its orientation. High-order resonances ( $f>2$) may be common in known tightly packed 
exoplanet systems where
planet's eccentricity may be excited by gravitational interaction with other planets \citep[e.g.,][]{gj}. 
Surviving exoplanets orbiting white dwarfs may also be captured into this kind of resonance, but its condition in the
$\sigma$--$e$ space remains to be investigated.

\begin{figure}
\plotone{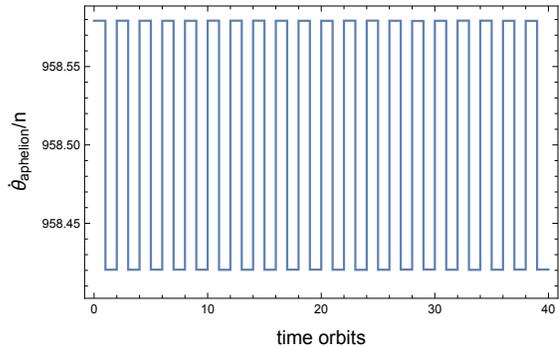}
\caption{Aphelion rotation velocity in the ($958.5\pm0.079)n$ {\it jumping} spin-orbit resonance of 2006 HY51 obtained from high-accuracy numerical integration of the ODE of motion.
\label{jum.fig} }
\end{figure}

The second new kind, called jumping resonance, is characterized by a nearly constant aphelion orientation $\theta$
modulo $\pi$, which 
is close to $\pi/2$ (sidewise), and aphelion velocity jumping between two quantum states for each pair of orbits, 
separated from an exact commensurability by a finite value. The possibility of a stable spin-orbit resonance with 
non-commensurate
spin rate has never been proposed, even theoretically. One example is the resonance at $\omega = 958.57923 n$ alternating with $958.42041 n$, shown in Fig. \ref{jum.fig}. As in the previous graph, the stepwise broken
line is shown only to help the eye to visualize the alternating aphelion spin rate. The actual integrated curve is close to
this broken line everywhere except the short time spans of a few days around perihelion passages, where the spin
rate undergoes powerful variations much greater in amplitude then the range of this plot. The aphelion orientation angle
(not presented for brevity)
shows a random process-like variation within a very narrow range without any signs of a periodic libration-like
modulation. From our massive simulations of chaotic trajectories with random initial conditions, this non-commensurate 
resonance appears to be more rare (i.e., has a lower probability of capture) than 
the regular circulation resonance possibly because of the narrow range in $\theta$. The offset of $\omega$ from the commensurate
fraction causes the asteroid to enter the perihelion at a slight tilt of its longest axis to the direction to the Sun.
The tilt causes an asymmetric impulse driving $\omega$ to the other quantum state, resulting in a tilt with the opposite
sign at the next perihelion passage. Amazingly, this jumping equilibrium is stable, in that a small perturbation in either
parameter does not result in a larger change of aphelion parameters.

\section{Inroads into the full 3D case}
\label{3d.sec}
Full 3D simulations of Euler's equations are in order to check the stability of high-spin resonances in the presence of obliquity wobble. The challenges are more daunting in 3D because of the immense parameter space that has to be mapped to locate
and measure the resonances. Beside the 6 initial condition parameters (orientation angle and spin rate for each axis),
the results are quite sensitive to eccentricity, orbital frequency, and relative distribution of the moments of inertia.
The three Euler's equations of rotation including Coriolis accelerations should be solved simultaneously as a single
system of ODEs. This dramatically raises the computing cost of simulations.
Our fast chaotic trajectory simulation method cannot be used, and other numerical methods have to be exploited to
distinguish chaotic trajectories from deterministic states.

\begin{figure}
\plottwo{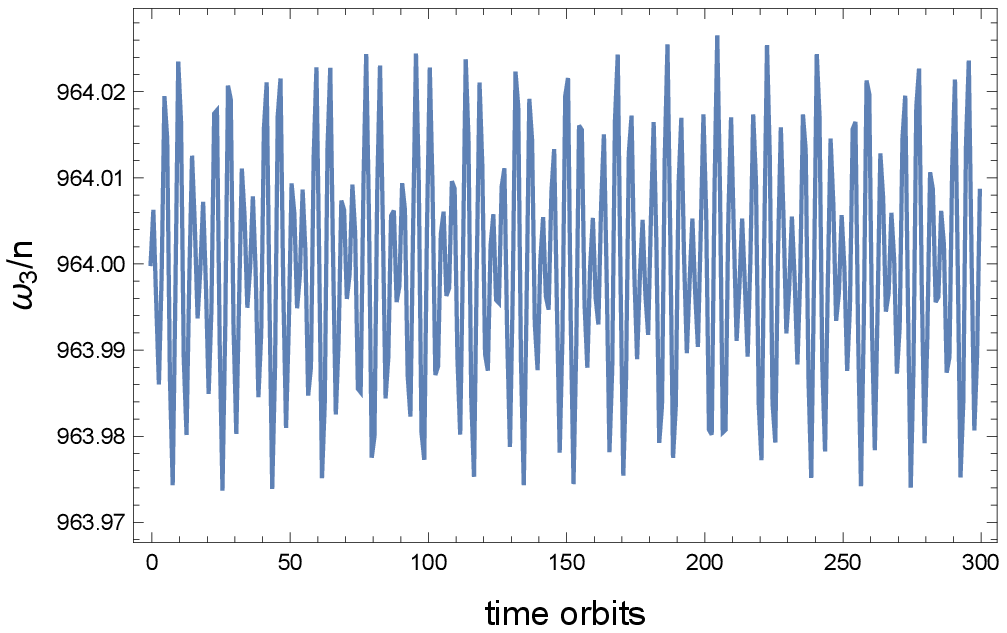}{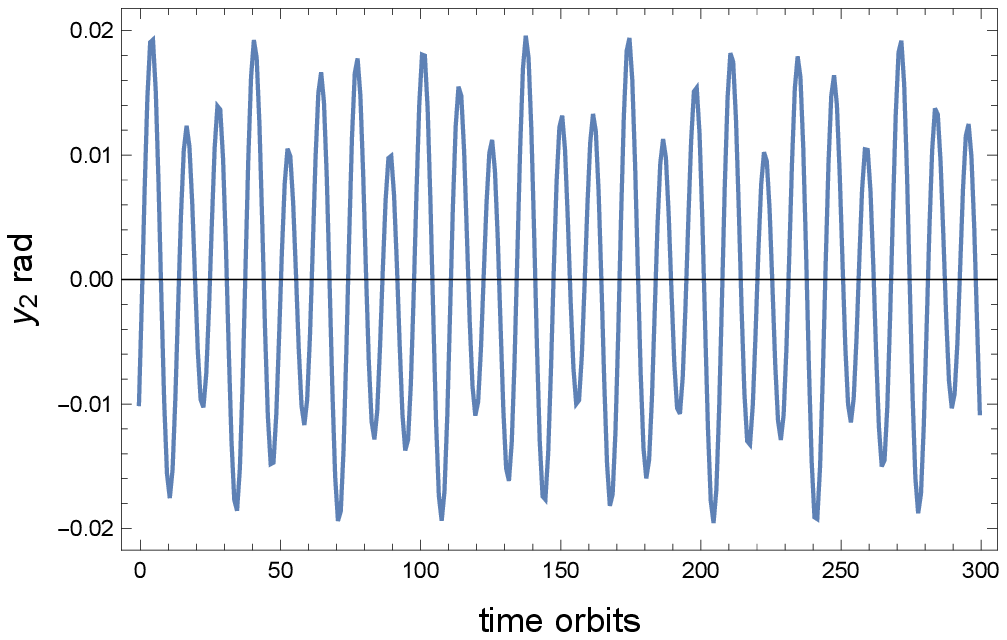}
\caption{Aphelion rotation velocity (left) and pitch angle (right) in the $964:1$ spin-orbit resonance of 2006 HY51 
obtained from high-accuracy numerical integration of the 3D Euler's equations of motion.
\label{3d.fig} }
\end{figure}

We performed limited 3D simulations to verify some of the resonances described in this paper. Fig. \ref{3d.fig}
shows the results of a full-scale numerical integration for 2006 HY 51 with stiffness switching for 300 orbits (1251
years). The initial conditions for the starting point at aphelion are:
$y_1=-0.01$ rad, $y_2=-0.01$ rad, $y_3=\pi$ rad, $\omega_1=0.01\,n$, $\omega_2=-0.005\,n$, $\omega_3=964\,n$.
The triaxial inertia coefficients are $(B - A)/C=0.1611$, $(C - A)/B=0.3423$, $(C - B)/A=0.1918$.
The left panel displays the evolution of normalized spin rate around the principal axis of the largest inertia
moment at aphelia. The right plot shows the evolution of one of the orientation angles (pitch) also at aphelia.
Both functions display a complex pattern of free libration with amplitudes of up to $0.024\,n$ and 0.02 rad,
respectively. The important conclusion is that the $964:1$ spin orbit resonance is real and long-term stable in the absence
of external perturbations. However, if we try to significantly increase the initial deviations of spin rates from
these values, we obtain trajectories that show initially small perturbations, which exponentially grow in amplitude
as the object literally begins to spin out of control in about 200 orbits. This implies that the investigated
resonance is rather narrow in terms of initial orientation angles and spin. Consequently, the characteristic times of 
capture and chaotic evolution may become longer in 3D, and the life expectation inside the resonance shorter. It remains to be
seen if other equilibria exist at $\omega_3=964\,n$ and nonzero commensurate spin rates in the other two dimensions.

\section{Conclusions}
\label{con.sec}

By simulating chaotic rotation of high-eccentricity asteroids with our fast tuple-generating method for ~1 Gyr time intervals, we discovered the existence of high spin-orbit resonances.
These stable resonances may serve as protection against rotational break up at higher eccentricity, which needs to be investigated separately.
Apart from the regular circulation resonances maintaining both aphelion velocity and orientation angle nearly constant, we discovered two new kinds of spin-orbit resonances: a switching resonance alternating the orientation angle, and a non-commensurate jumping resonance alternating the aphelion velocity between two quantum states. All three kinds are
intrinsically stable, but the disposition of specific commensurabilities and the width of spin-orbit resonances
are both eccentricity- and triaxiality-dependent, which is demonstrated for the regular 964:1 resonance
using a chaos detection method.

The characteristic times of capture with 2006 HY51 parameters are well below 100 Myr.
Therefore, 2006 HY51 may be captured in one of such resonances, which would be interesting to verify by observation.
Rotation period can be inferred from systematic photometric observations, and two campaigns bracketing a perihelion
conjunction could reveal if the asteroid maintained a resonance velocity and its value. On the other hand,
if 2006 HY51 happens to be in the chaotic state of rotation, a significant and measurable update could be observed.
Because of the narrowness of these equilibrium states, a moderate external perturbation (from an inner planet, for example) may extract the asteroid and trigger another chaotic walk for millions of years.

\section*{Acknowledgments}

DV gratefully acknowledges the support of the STFC via an Ernest Rutherford Fellowship (grant ST/P003850/1).
We used Julia \citep{rac} to implement our chaotic rotation simulation method, which is available as open source
software via github.

\end{document}